\documentclass[10pt, conference, compsocconf]{IEEEtran}
\usepackage[table]{xcolor}
\usepackage{graphicx}

\usepackage{multirow}
\usepackage{verbatim}
\usepackage{fmtcount}
\usepackage{threeparttable}
\usepackage[font=footnotesize]{caption}
\usepackage{lipsum}
\usepackage{tabularx,colortbl}
\usepackage{subfig}
\usepackage{rotating}
\usepackage{adjustbox}
\usepackage{url}
\usepackage{float}
\usepackage{pdflscape,array,booktabs}
\usepackage{booktabs}
\usepackage{amsthm}
\usepackage{amsmath}
\usepackage{multicol}

\DeclareCaptionType{copyrightbox}

\usepackage{listings,multicol}
 \usepackage{courier}

\definecolor{darkgray}{rgb}{0.95,0.95,0.95}
\DeclareGraphicsExtensions{.pdf,.png,.jpg}

\definecolor{maroon}{rgb}{0.5,0,0}
\definecolor{darkgreen}{rgb}{0,0.5,0}

\lstdefinelanguage{XML}
{
  basicstyle=\ttfamily,
  morestring=[s]{"}{"},
  morecomment=[s]{?}{?},
  morecomment=[s]{!--}{--},
  commentstyle=\color{darkgreen},
  moredelim=[s][\color{black}]{>}{<},
  moredelim=[s][\color{red}]{\ }{=},
  stringstyle=\color{blue},
  identifierstyle=\color{maroon}
}

\usepackage{cite}

\ifCLASSINFOpdf
\else
\fi
\usepackage{url}

\usepackage{xcolor}
\usepackage{todonotes}

\IEEEoverridecommandlockouts

\makeatletter
\def\@outputdblcol{%
  \if@firstcolumn
    \global\@firstcolumnfalse
    \global\setbox\@leftcolumn\copy\@outputbox
    \splitmaxdepth\maxdimen
    \vbadness\maxdimen

    \setbox\@outputbox\vbox{\unvbox\@outputbox\unskip}%
    \setbox\@outputbox\vsplit\@outputbox to\maxdimen
    \toks@\expandafter{\topmark}%
    \xdef\@firstcoltopmark{\the\toks@}%
    \toks@\expandafter{\splitfirstmark}%
    \xdef\@firstcolfirstmark{\the\toks@}%
    \ifx\@firstcolfirstmark\@empty
      \global\let\@setmarks\relax
    \else
      \gdef\@setmarks{%
        \let\firstmark\@firstcolfirstmark
        \let\topmark\@firstcoltopmark}%
    \fi
  \else
    \global\@firstcolumntrue
    \setbox\@outputbox\vbox{%
     \hb@xt@\textwidth{%
        \hb@xt@\columnwidth{\box\@leftcolumn \hss}%
        \hfil
        \vrule \@width\columnseprule
        \hfil
       \hb@xt@\columnwidth{\box\@outputbox \hss}}}%
  \@combinedblfloats
    \@setmarks
    \@outputpage
    \begingroup
      \@dblfloatplacement
      \@startdblcolumn
      \@whilesw\if@fcolmade \fi{\@outputpage\@startdblcolumn}%
    \endgroup
  \fi}
\makeatother
\begin{document}
\pagestyle{plain}

%
\title{\LARGE A Collaborative Model for Improving Information Sharing among Cancer Care Groups using Software Engineering Principles}


\author{\IEEEauthorblockN{Davis Byamugisha}
\IEEEauthorblockA{Mbarara University of Science \\\newline and Technology \\\newline davisb256@gmail.com}
\and
\IEEEauthorblockN{Francis Kamuganga}
\IEEEauthorblockA{Mbarara University of Science \\\newline and Technology \\\newline fkamg@must.ac.ug}
\and
\IEEEauthorblockN{Adones Rukundo}
\IEEEauthorblockA{Mbarara University of Science \\\newline and Technology \\\newline adones@must.ac.ug}
\and
\IEEEauthorblockN{John Businge}
\IEEEauthorblockA{University of Nevada \\\newline
john.businge@unlv.edu}
}


%



\maketitle

\begin{abstract}
Effective treatment of cancer requires early diagnosis which involves the patient’s awareness of the early signs and symptoms, leading to a consultation with a health provider, who would then promptly refer the patient for confirmation of the diagnosis and thereafter treatment. However, this is not always the case because of delays arising from limited skilled manpower and health information management systems that are neither integrated nor organized in their design hence leading to information gap among care groups. Existing methods focus on using accumulated data to support decision-making, enhancing the sharing of secondary data while others exclude some critical stakeholders like patient caretakers and administrators thus, leaving an information gap that creates delays and miscommunication during case management. We however notice some similarities between cancer treatment and software engineering information management especially when progress history needs to be maintained (versioning). 

We analyse the similarities and propose a model for information sharing among cancer care groups using the software engineering principles approach. We model for reducing delays and improving coordination among care groups in cancer case management. Model design was guided by software engineering principles adopted in GitHub version control system for bug fixing in open-source code projects. Any-Logic simulation software was used to mimic the model realism in a virtual environment. Results show that bug resolution principles from software engineering and GitHub version control system can be adopted to coordinate collaboration and information sharing among care groups in a cancer case management environment while involving all stakeholders to improve care treatment outcomes, ensure early diagnosis and increase patient’s survival chances.


\footnote{This work has been funded by the government of Uganda through Mbarara University of Science and Technology Directorate of Research and Graduate Training under grant No. DRGT/SG/FY23-24/R4/T1P4.}
\end{abstract}

\section{Introduction}
\label{sec:intro}
Globally, cancer incidence and mortality have been increasing rapidly in recent years according to reports produced by the International Agency for Research on Cancer (IARC) \cite{bray2018global}. Cancer is now a leading cause of death worldwide and according to the Word Health Organisation(WHO), 66\% of all cancer-related deaths occur in low and middle-income countries, where resources available for prevention, diagnosis, and treatment of cancer are limited or even nonexistent \cite{10665-254500}. The advanced human expertise and technical infrastructure needed for optimum diagnosis, treatment, and management of cancer cases is already difficult enough in high-income countries and becomes especially challenging in low and middle-income countries \cite{doi:10.1146/annurev-pathol-042320-034052}.\par
Late-stage presentation and inaccessible diagnosis and treatment are prevalent in low and middle-income countries where cancer fatality rates are much higher according to WHO, the number of cancer cases is expected to almost double by 2040 yet many cancers have a high chance of cure if diagnosed early and treated adequately \cite{10665-254500}. The repercussions of delays in care and advanced cancer are severe because the likelihood of death and disability from cancer increases significantly as cancer progresses. It is therefore critical to identify barriers to timely diagnosis and treatment and to implement novel interventions that resolve these barriers thus improving cancer management \cite{doi:10.1200/JGO.18.00200}.\par
For effective treatment of cancer, early diagnosis is important. Early diagnosis often involves the patient’s awareness of the early signs and symptoms, leading to a consultation with a health provider, who would then promptly refer the patient for confirmation of the diagnosis and thereafter treatment. However, this is not always the case because of delays that may lead to screening happening outside recommended intervals, confusion among the care providers, and ambiguities about the patient’s role in care \cite{Taplin2015}.
\par
Coordination amongst providers during cancer care plays a significant role, however, it is challenging due to its complexity \cite{baker2001crossing}. The care of most cancer patients frequently involves multi-modal cancer therapies, the treatment of other medical conditions apart from cancer itself, and multiple health services, providers, and sites of care \cite{levit2013delivering, baker2001crossing}. The complexity of cancer care coordination is further aggravated by delivery systems that are neither integrated nor organised in their design.\par
According to the WHO, delays in timely diagnosis and accessing treatment can occur at multiple steps along the patient journey which could be (i) delays in awareness of symptoms, (ii) delays in clinical evaluation, and diagnosis and (iii) delays in accessing treatment. These steps are independent of each other and there is often no shared information about the cancer patient or their condition across the steps.\par
Due to the information gap in cancer case management, several studies have been conducted to find a lasting solution for example the National Cancer Institute (NCI) of USA \cite{Anu:2022} and the International Cancer Clinic recommended a computerised platform to enhance information exchange during case trials and management \cite{trimble2009improving} however, the proposed platform was not implemented.\par
Another study was carried out by Tamposis et al. \cite{tamposis2022pcaguard}, in their investigation on current aspects of implementation and architectural design of health care systems to develop a software platform to support prostate cancer management. Their findings show that conducting patient follow-ups using technology can be effective and productive. However, the developed platform focused on using accumulated data to estimate the risks of prostate cancer detection hence cancer case management collaboration wasn’t catered for in this study. \par
Stuart et al. \cite{watt2013clinical} also proposed a software application to track patients, sample acquisition, genome results, and reports during cancer management. The system integrates modules for reading, storing, and analysing data to support decision-making. Also, Tamposis et al. \cite{7015996} proposed a software platform to support cervical cancer prediction and management. In their design, dependency inversion software engineering principle \cite{o2018application} was applied to study patient medical history to support cervical cancer prediction. Though much has been achieved, the current studies focus on sharing secondary data to support decision-making. This is equally important though there was a need for a method to support case collaboration in active management. Although there are no documented interventions carried out in Uganda about coordinated cancer case management, several studies have been conducted assessing the need for improved case management as well as suggesting possible solutions for improving the situation. For example, Amos et al \cite{mwaka2016cancer} assessed how cancer cases are handled at Mulago national hospital. In their findings it was reported that there are few specialised cancer facilities and cancer specialists in Uganda thus, they are subjected to intensive workload with no automated systems to assist in case management\cite{mwaka2016cancer}.
\par
Therefore, to address the challenge of the information gap among care groups, in this study, we proposed a collaborative model for reducing delays and improving coordination among care groups in cancer case management.
\section{Literature review}
McCorkle et al. \cite{mccorkle2015advanced} developed a model for monitoring patients’ status, managing symptoms, offering training to the patient and caregivers, coordinating care among care groups, responding to patient’s immediate family members, and executing complex care procedures. Additionally, the model supports collaborations among care providers such as laboratory technicians. In this study, researchers aimed to assess the effect of multidisciplinary coordinated intervention on outcomes with patients. The study focused on late-stage cancer patients. Despite the study achievements, coordination is made through weekly phone calls and in-person contact which creates delays in the case management coordination process since phone calls and in-person contact does not guarantee timely information availability.\par
Wulff et al. \cite{wulff2012randomised} implemented a method for managing colorectal cancer cases in Denmark. Using this method, a case manager can connect with the patient, and assess the patient’s biopsychosocial status, perform screening of barriers to optimal care related to coordination and awareness of the care plan. The main objective of this study was to analyze the success of hospital-based cancer care management in terms of patient-reported outcomes. In this method, face-to-face, telephone, and electronic letters are used to effect communication between the care manager and the patient. This method is limited since it does not involve patient’s immediate family care providers, laboratory technicians plus other care teams. This affects case management process since there is a gap in information flow among care teams.\par
Scherz et al. \cite{scherz2017case} implemented an intervention for cancer case management in Switzerland using electronic messages and telephone calls to coordinate communication between the care manager and the patient. The method supports needs assessment, generates action plans, provides information on available services and therapies, and schedules appointments.
Besides eliminating other care teams like patient's immediate family members which creates an information gap, electronic messages and telephone calls can lead to increased delays in case management especially if a patient misses a call from the care manager.\par
Ozcelik et al. \cite{ozcelik2014examining} proposed a method for coordinating the management of case symptoms, extending social support to the patient, enabling family counseling services, and psycho-social stress management. The study aimed at investigating improvement in symptoms, quality of life, and family and patient satisfaction with care. Whereas counseling is done during hospital admission, appointment follow-ups are scheduled and reminders are triggered for the upcoming appointment. This method mainly focused on counseling and family-social care teams hence ignoring the case manager and other health teams who are vital in the case management process. The existing information gap among care teams creates more delays which affect the entire case management process.\par
In a study by Kok Swee Sim et al. \cite{sim2014computerized}, a computerized database management system for managing breast cancer was developed. The system supports storing and retrieving patient data, manipulating data records, and appointment scheduling. Besides, the system automated data analytics for cancer risk factors assessment. MySQL database and Microsoft Access tools were used to design the logical structure of the database. Despite the success stories of the system in Malaysia, it does not facilitate coordination of the cancer case management process i.e., the system only stores patients' records and cancer risk factors assessment but does not facilitate information sharing.\par
 Pei-Yi Lee and Tsue-Rung Chang \cite{lee2015application} developed an integrative information system to improve the effectiveness and quality of data management in cancer case management systems. The system design was guided by suggestions and in-depth and detailed revision of comments from various case managers. The system has an embedded framework that facilitates the design of case management flow and case management planning. System design functionalities are limited to the medical team thus, neglecting the patient family caregiver, counselors, and psycho-social team despite their contribution towards the patient’s adherence to medication.\par
From the literature, it is observed, that the developed platform focuses on using accumulated data to support decision-making, enhancing the sharing of secondary data while others exclude some stakeholders like patients’ caretakers and administrators thus, leaving an information gap that creates delays and miscommunication during case management. Thus there was a need to adopt the software engineering principle used in the GitHub version control system to simulate a model for enhancing information sharing and support collaboration among groups during cancer case management. 
\section{Methodology}
\subsection{Model requirement selection}
The selection of model requirements was guided by the application of software engineering principles, specifically those employed in the process of software bug fixing. This process involves several key steps, including:\par
\textbf{Core team establishment:} From the Pareto principle \cite{dunford2014pareto}, it is assumed that 80\% of the output is produced by 20\% of the collaborators i.e., the core team. Thus, the proposed model is assumed to incorporate a module for establishing a core team led by the case manager. A case manager is a healthcare professional who serves as a patient advocate to support, guide, and coordinate care for the patient, families, and caregivers as they navigate their health and wellness journeys \cite{kanter1989clinical} \par
\textbf{Information management:} During case management, a lot of information is gathered at different stages by different teams thus, information management is necessary to track its source, integrity, and consumption. The designed model adopts a GitHub Version Control System (VCS) \cite{blischak2016quick} to monitor the activities of all collaborators and perform information validation before integration and consumption.
\textbf{Inclusion data access:} Successful case management requires timely information access by all involved stakeholders, cancer case management involves various teams including administrators, doctors, nurses, psycho-social, and close family caregivers. Each of these groups/stakeholders plays different roles and need access to different information. To control information access, the designed model applies the discretionary access control principle to grant information access rights based on the identity of the subject.\par
\textbf{Case monitoring:} The Cancer case monitoring process involves various activities including; medication coordination, medical records transfer, laboratory results follow-up, communication coordination, treatment side effects monitoring, and coordinating medication appointments \cite{silbermann2013multidisciplinary}. The activities performed are categorized as diagnosis, information gathering, iterating potential solutions, and treatment assessment stages of case management \cite{hudon2019characteristics}. The model is designed to take into consideration these stages.\par
The above-mentioned requirements were used to determine model blocks and information flow among the entities.
\subsection{Model design and architecture}
The phase of model design started after requirements analysis, using the principles of software engineering to organize the architecture of the system. Modeling the process of cancer case management as a sequence of discrete events, the model was designed as a discrete-event simulation (DES) system \cite{forbus2022discrete}. For example, patient referrals, diagnostic testing, therapy sessions, and follow-up appointments are all associated with specific events that take place during the care process.
Interconnected components representing various care groups, including radiologists, surgeons, oncologists, and nursing teams, comprised the modular architecture. The decision-making procedures, group interactions, and patient case management of each module were all intended to mimic the actions of the corresponding care group. Model modifications were made flexible and future extensions may be scaled.
\subsection{Simulation Modeling}
The designed model was implemented using a simulation using DES techniques \cite{forbus2022discrete}. The simulation environment was developed in Anylogic a specialized simulation software platform \cite{enwiki:1081908536}, chosen for its capability to handle complex healthcare processes and its support for discrete-event simulations. The Key steps observed in this simulation include;
\begin{enumerate}
    \item \textbf{Event scheduling:} The simulation engine was programmed to handle the dynamic scheduling of events based on real-world data, such as patient arrival times, treatment durations, and care group availability. The event scheduler was designed to prioritize critical events and manage potential conflicts, such as double-booked appointments or resource shortages.
    \item \textbf{Process flow definition:} The flow of patients through the healthcare system was modeled using state transitions, where each state represented a stage in the care process. The transitions between states were governed by probabilistic rules derived from empirical data, capturing the variability and uncertainty inherent in cancer care management
    \item \textbf{Resource management:} Within the simulation, resources such as medical personnel, facilities, and equipment were represented as finite entities. Algorithms that attempted to reduce delays and enhance overall coordination were used to optimize the distribution of these resources across different jobs. Real-time tracking of resource use by the simulation made it possible to spot inefficiencies and bottlenecks.
    \item \textbf{Interaction and coordination mechanisms:} Coordination guidelines and communication protocols were included in the simulation to simulate the interactions between various care groups. To ensure that the model appropriately reflected the collaborative character of cancer case management, these mechanisms were created to mirror real-world coordination activities, such as interdisciplinary meetings and electronic health record (EHR) exchanges.
\end{enumerate}
\subsection{Model testing}
Following its initial deployment, the simulation model was rigorously validated and tested. To make sure the simulated outcomes were feasible and realistic, the validation approach comprised comparing the model's outputs with historical data from real cancer case management scenarios. Sensitivity analysis was also carried out to evaluate how changing important parameters—like patient arrival rates and resource availability—would affect the model's functionality. \par
During the testing phase, the model was utilized to simulate various coordination strategies and how they affected the amount of delay. These scenarios guided recommendations for workable treatments by offering insights into possible enhancements in care coordination.
\section{Results}
\subsection{A model for improving information sharing in cancer case management}
The designed model has two layers; i.e. i) Core Team Collaboration and ii) Information Access layers. The collaboration layer offers collaboration privileges to the core team which is directly involved in administering medication. Whereas the information access layer implements an inclusion of information access to other stakeholders with access rights. The model operations are described in seven steps as illustrated in Figure 1.2.
\subsection{Model description}
\textbf{Case enrolment (1):} This is the first phase and it involves patient/case enrolment, at this level the case manager registers case details including the patient’s bio-data, any available medical history, caretaker’s details plus any other information deemed necessary for case management. This can be equated to the GitHub version control system where an issue is identified and submitted to the repository with a title and full description with all relevant details.\par
\textbf{Core team setup (2):} Once a case is established, the case manager establishes a core team including the nurses, lab technicians and any other allied health staff such as diagnostic medical sonographers, dietitians, medical technologists, occupational therapists, physical therapists, radiographers, respiratory therapists, speech language pathologists etc.\par 
\textbf{Open collaboration (3):} The established team is notified through a collaboration request shared by the case manager. Collaboration requests include the collaborator’s team, the role to be played, case management schedules, and prior information to assist in executing the assigned task.\par
\textbf{Collaborator input (4):} During case execution, every collaborator is entitled to provide timely updates for the assigned role. Thus, at step four GitHub version control system is adopted to track collaborator inputs, and support input validation by the case manager. In a situation where irregularities are noticed, a collaborator is notified and requested to investigate the limitations and update the input.\par
\textbf{Approve input (5,7):} This is the second last step where the case manager approves the collaborator's inputs and this is followed by merging (updating) information for consumption by the entire collaboration team and other stakeholders.\par
\textbf{Information access (7):} This layer ensures only stakeholders with “access rights” access patient’s data, among the stakeholders include; (i) the patient and patient's caretaker(s) who view(s) laboratory results, medical treatment plans, medical team details, and scheduled appointments, (ii) the administrator – these are hospital administrative staff members with information view rights to enable case progress tracking and (iii) the psycho-social team providing counselling services for psychological patient empowerment to ensure medical adherence.
\begin{figure*}[htp]
     \centering
     \includegraphics[width=17cm, height=10cm]{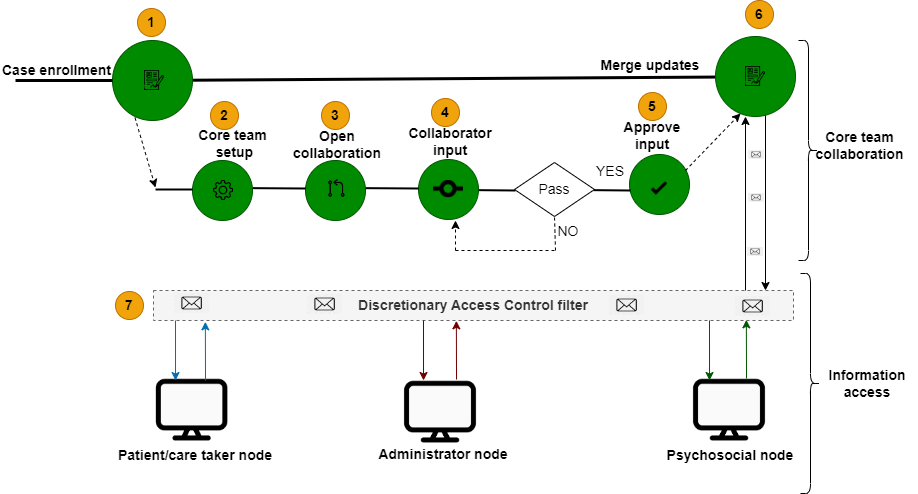}
     \caption{\textit{A model for improving information sharing and reducing delays in cancer case management }}
\end{figure*}
\section{Discusssion}
Several studies have been conducted where several distributed medical team collaboration platforms have been proposed, for example, Zhan Zhang et al \cite{zhang2017constructing}.
These methods mainly focused on collaboration among medical teams thus, neglecting other stakeholders like the patient's family members, and the psycho-social team which affects case management processes yet these groups play significant roles during case management. The designed model caters for all stakeholders involved in cancer case management. Furthermore, the designed model accounts for the overall general medical case management processes, and thus it could even be adopted in managing other long-term medical complications such as Diabetes, HIV-AIDS, and Hypertension.\par
The simulated model experiment is based on GitHub version control system to mimic information flow in a cancer case management environment. The simulation test operates under several assumptions; a well-established technology infrastructure is in place, no user response delays, with an assumed latency of 0.25 minutes for receiving feedback from collaborators, and input validation latency is set at 1440 seconds. By executing the model within a controlled virtual environment, the movement of data packets is carefully monitored, allowing for a detailed assessment of the model's performance under these conditions.\par
From the simulation experiment, patient information is observed moving from multiple sources to the receiving ends while accounting for the collaboration protocols, and timely, restricted, and remote access to the patients' information hence mimicking the operation of the proposed model based on the GitHub version control system. Based on these findings, the researcher believes that if the proposed model is deployed in a real environment while observing the procedure, delays related to information sharing will be reduced hence improving timely case management in a cancer care management environment.

\section{Conclusion and further work}
The investigation revealed that collaboration tools are potential solutions for reducing delays involved in cancer case management, thus the designed model has demonstrated the potential infrastructure towards realising technology adoption in cancer case management. On this basis, future research should examine the impact of technology adoption on improving care group productivity and care treatment outcomes.\par
Whereas the study assessed the possibility of adopting GitHub software engineering principles for bug fixing in cancer case management, the study was limited to open-source code projects where collaborators do not wait for collaboration invitations. However, the designed model assumes a collaboration request being initiated by the lead collaborator(case manager) hence future studies should assess the applicability of GitHub software engineering principles used by collaborators on private projects. Additionally, the designed model was simulated with assumed parameters, hence there is a need to assess the model performance in a real-world environment by integrating it in existing medical records management systems.  


\bibliographystyle{elsarticle-num}
\bibliography{biblio}

@article{bray2018global,
  title={Global cancer statistics 2018: GLOBOCAN estimates of incidence and mortality worldwide for 36 cancers in 185 countries},
  author={Bray, Freddie and Ferlay, Jacques and Soerjomataram, Isabelle and Siegel, Rebecca L and Torre, Lindsey A and Jemal, Ahmedin},
  journal={CA: a cancer journal for clinicians},
  volume={68},
  number={6},
  pages={394--424},
  year={2018},
  publisher={Wiley Online Library}
}

@BOOK{10665-254500,
	author = {World Health Organization},
	title = {Guide to cancer early diagnosis},
	year = {2017},
	pages = {48 p.},
	publisher = {World Health Organization},
	type = {Publications}
}

@article{doi:10.1146/annurev-pathol-042320-034052,
author = {Radich, Jerald P. and Briercheck, Edward and Chiu, Daniel T. and Menon, Manoj P. and Sala Torra, Olga and Yeung, Cecilia C.S. and Warren, Edus H.},
title = {Precision Medicine in Low- and Middle-Income Countries},
journal = {Annual Review of Pathology: Mechanisms of Disease},
volume = {17},
number = {1},
pages = {387-402},
year = {2022}
}

@article{doi:10.1200/JGO.18.00200,
author = {Shah, Shailja C. and Kayamba, Violet and Peek, Richard M. and Heimburger, Douglas},
title = {Cancer Control in Low- and Middle-Income Countries: Is It Time to Consider Screening?},
journal = {Journal of Global Oncology},
pages = {1-8},
year = {2019}
}

@article{Taplin2015,
   author = {Stephen H Taplin and Sallie Weaver and Veronica Chollette and Lawrence B Marks and Andrew Jacobs and Gordon Schiff and Carrie T Stricker and Suanna S Bruinooge and Eduardo Salas},
   issue = {3},
   journal = {Journal of oncology practice},
   pages = {231-238},
   publisher = {American Society of Clinical Oncology Alexandria, VA},
   title = {Teams and teamwork during a cancer diagnosis: interdependency within and between teams},
   volume = {11},
   year = {2015},
}

@book{baker2001crossing,
  title={Crossing the quality chasm: a new health system for the 21st century},
  author={Baker, Alastair},
  volume={323},
  year={2001},
  publisher={British Medical Journal Publishing Group}
}

@article{tamposis2022pcaguard,
  title={PCaGuard: A Software Platform to Support Optimal Management of Prostate Cancer},
  author={Tamposis, Ioannis and Tsougos, Ioannis and Karatzas, Anastasios and Vassiou, Katerina and Vlychou, Marianna and Tzortzis, Vasileios},
  journal={Applied Clinical Informatics},
  volume={13},
  number={01},
  pages={091--099},
  year={2022},
  publisher={Georg Thieme Verlag KG}
}

@article{watt2013clinical,
  title={Clinical genomics information management software linking cancer genome sequence and clinical decisions},
  author={Watt, Stuart and Jiao, Wei and Brown, Andrew MK and Petrocelli, Teresa and Tran, Ben and Zhang, Tong and McPherson, John D and Kamel-Reid, Suzanne and Bedard, Philippe L and Onetto, Nicole and others},
  journal={Genomics},
  volume={102},
  number={3},
  pages={140--147},
  year={2013},
  publisher={Elsevier}
}

@INPROCEEDINGS{7015996,
  author={Tamposis, Ioannis and Iordanidis, Evripidis and Tzortzis, Leonidas and Bountris, Panagiotis and Haritou, Maria and Koutsouris, Dimitrios and Pouliakis, Abraham and Karakitsos, Petros},
  booktitle={2014 4th International Conference on Wireless Mobile Communication and Healthcare}, 
  title={HPVGuard: A software platform to support management and prognosis of cervical cancer}, 
  year={2014},
  volume={},
  number={},
  pages={401-405},
  }

@inproceedings{o2018application,
  title={Application of the Dependency Inversion Principle to Multidisciplinary Software Development},
  author={O'Connell, Matthew and Druyor, Cameron and Thompson, Kyle B and Jacobson, Kevin and Anderson, William K and Nielsen, Eric J and Carlson, Jan-Rene{\'e} and Park, Michael A and Jones, William T and Biedron, Robert and others},
  booktitle={2018 Fluid Dynamics Conference},
  pages={3856},
  year={2018}
}

@book{levit2013delivering,
  title={Delivering high-quality cancer care: charting a new course for a system in crisis},
  author={Levit, Laura A and Balogh, Erin and Nass, Sharyl J and Ganz, Patricia and others},
  year={2013},
  publisher={National Academies Press Washington, DC}
}

@inproceedings{zhang2017constructing,
  title={Constructing common information spaces across distributed emergency medical teams},
  author={Zhang, Zhan and Sarcevic, Aleksandra and Bossen, Claus},
  booktitle={Proceedings of the 2017 ACM Conference on Computer Supported Cooperative Work and Social Computing},
  pages={934--947},
  year={2017}
}

@misc{ enwiki:1081908536,
    author = "{Wikipedia contributors}",
    title = "AnyLogic --- {Wikipedia}{,} The Free Encyclopedia",
    year = "2022",
    note = "[Online; accessed 13-April-2022]"
  }

@article{blischak2016quick,
  title={A quick introduction to version control with Git and GitHub},
  author={Blischak, John D and Davenport, Emily R and Wilson, Greg},
  journal={PLoS computational biology},
  volume={12},
  number={1},
  pages={e1004668},
  year={2016},
  publisher={Public Library of Science}
}

@MISC{Anu:2022,
author = {USA NCI},
title = {Clinical Trials Information for Patients and Caregivers},
month = {August},
year = {2022},
howpublished={\url{https://www.cancer.gov/about-cancer/treatment/clinical-trials}}
}

@article{trimble2009improving,
  title={Improving cancer outcomes through international collaboration in academic cancer treatment trials},
  author={Trimble, Edward L and Abrams, Jeffrey S and Meyer, Ralph M and Calvo, Fabien and Cazap, Eduardo and Deye, James and Eisenhauer, Elizabeth and Fitzgerald, Thomas J and Lacombe, Denis and Parmar, Max and others},
  journal={Journal of clinical oncology},
  volume={27},
  number={30},
  pages={5109},
  year={2009},
  publisher={American Society of Clinical Oncology}
}

@article{silbermann2013multidisciplinary,
  title={Multidisciplinary care team for cancer patients and its implementation in several Middle Eastern countries},
  author={Silbermann, M and Pitsillides, B and Al-Alfi, N and Omran, S and Al-Jabri, K and Elshamy, K and Ghrayeb, I and Livneh, J and Daher, M and Charalambous, H and others},
  journal={Annals of oncology},
  volume={24},
  pages={vii41--vii47},
  year={2013},
  publisher={Elsevier}
}

@article{mccorkle2015advanced,
  title={An advanced practice nurse coordinated multidisciplinary intervention for patients with late-stage cancer: a cluster randomized trial},
  author={McCorkle, Ruth and Jeon, Sangchoon and Ercolano, Elizabeth and Lazenby, Mark and Reid, Amanda and Davies, Marianne and Viveiros, Diane and Gettinger, Scott},
  journal={Journal of palliative medicine},
  volume={18},
  number={11},
  pages={962--969},
  year={2015},
  publisher={Mary Ann Liebert, Inc. 140 Huguenot Street, 3rd Floor New Rochelle, NY 10801 USA}
}

@article{wulff2012randomised,
  title={A randomised controlled trial of hospital-based case management to improve colorectal cancer patients’ health-related quality of life and evaluations of care},
  author={Wulff, Christian Nielsen and Vedsted, Peter and S{\o}ndergaard, Jens},
  journal={BMJ open},
  volume={2},
  number={6},
  pages={e001481},
  year={2012},
  publisher={British Medical Journal Publishing Group}
}

@article{scherz2017case,
  title={Case management to increase quality of life after cancer treatment: a randomized controlled trial},
  author={Scherz, Nathalie and Bachmann-Mettler, Ir{\`e}ne and Chmiel, Corinne and Senn, Oliver and Boss, Nathalie and Bardheci, Katarina and Rosemann, Thomas},
  journal={BMC cancer},
  volume={17},
  number={1},
  pages={1--8},
  year={2017},
  publisher={Springer}
}

@article{ozcelik2014examining,
  title={Examining the effect of the case management model on patient results in the palliative care of patients with cancer},
  author={Ozcelik, Hanife and Fadiloglu, Cicek and Karabulut, Bulent and Uyar, Meltem},
  journal={American Journal of Hospice and Palliative Medicine{\textregistered}},
  volume={31},
  number={6},
  pages={655--664},
  year={2014},
  publisher={SAGE Publications Sage CA: Los Angeles, CA}
}

@article{sim2014computerized,
  title={Computerized database management system for breast cancer patients},
  author={Sim, Kok Swee and Chong, Sze Siang and Tso, Chih Ping and Nia, Mohsen Esmaeili and Chong, Aun Kee and Abbas, Siti Fathimah},
  journal={SpringerPlus},
  volume={3},
  number={1},
  pages={1--16},
  year={2014},
  publisher={SpringerOpen}
}

@article{lee2015application,
  title={Application of integrative information system improves the quality and effectiveness of cancer case management},
  author={Lee, Pei-Yi and Chang, Tsue-Rung},
  journal={Journal of Multidisciplinary Healthcare},
  volume={8},
  pages={287},
  year={2015},
  publisher={Dove Press}
}

@article{hudon2019characteristics,
  title={Characteristics of case management in primary care associated with positive outcomes for frequent users of health care: a systematic review},
  author={Hudon, Catherine and Chouinard, Maud-Christine and Pluye, Pierre and El Sherif, Reem and Bush, Paula Louise and Rihoux, Beno{\^\i}t and Poitras, Marie-Eve and Lambert, Mireille and Zomahoun, Herv{\'e} Tchala Vignon and L{\'e}gar{\'e}, France},
  journal={The Annals of Family Medicine},
  volume={17},
  number={5},
  pages={448--458},
  year={2019},
  publisher={Annals Family Med}
}

@article{dunford2014pareto,
  title={The pareto principle},
  author={Dunford, Rosie and Su, Quanrong and Tamang, Ekraj},
  year={2014},
  journal={The Plymouth Student Scientist},
  publisher={University of Plymouth}
}

@incollection{mwaka2016cancer,
  title={Cancer Care in Countries in transition in Africa: the case of Uganda},
  author={Mwaka, Amos Deogratius and Wabinga, Henry and Garimoi, Christopher Orach},
  booktitle={Cancer Care in Countries and Societies in transition},
  pages={219--230},
  year={2016},
  publisher={Springer}
}

@article{kanter1989clinical,
  title={Clinical case management: Definition, principles, components},
  author={Kanter, Joel},
  journal={Psychiatric Services},
  volume={40},
  number={4},
  pages={361--368},
  year={1989},
  publisher={Am Psychiatric Assoc}
}

@article{forbus2022discrete,
  title={Discrete-event simulation in healthcare settings: a review},
  author={Forbus, John J and Berleant, Daniel},
  journal={Modelling},
  volume={3},
  number={4},
  pages={417--433},
  year={2022},
  publisher={MDPI}
}

\end{document}